\documentclass[a4paper]{jpconf}
\usepackage{amsmath}
\usepackage{amssymb}
\usepackage{amsfonts}
\usepackage{amscd}
\usepackage{bm}
\usepackage{stmaryrd}
\usepackage{bbold}
\usepackage{isomath}
\usepackage{xcolor}
\usepackage{graphicx}
\usepackage{theorem}
\usepackage{xspace}
\input xy
\xyoption{all}

\def\Z{\mathbb Z}
\def\R{\mathbb R}
\def\C{\mathbb C}
\def\H{\mathbb H}
\def\O{\mathbb O}

\def\demi{{\frac{1}{2}}}

\newcommand{\beq}{\begin{equation}}
\newcommand{\eeq}{\end{equation}}
\newcommand{\ba}{\begin{array}}
	\newcommand{\ea}{\end{array}}
\newcommand{\beqa}{\begin{eqnarray}}
\newcommand{\eeqa}{\end{eqnarray}}

\newtheorem{deff}[subsection]{Definition}

\newtheorem{prop}[subsection]{Proposition}

\def\R{\mathbb R}
\def\C{\mathbb C}
\def\H{\mathbb H}
\def\O{\mathbb O}

\def\JJ{\mathfrak J}
\def\demi{\frac{1}{2}}
\def\Aa{\mathcal A}
\begin{document}
\title{A remarkable dynamical symmetry\\ of the
Landau problem}

\author{Tekin Dereli$^1$, Philippe Nounahon$^2$, Todor Popov$^{3,4,5}$}

\address{$^1$Department of physics, Ko\c{c} University, 34450 Sariyer-\.{I}stanbul, Turkey}
\address{$^2$Institut de math\'ematique et de sciences physiques, B.P 613, Porto-Novo, B\'enin}
\address{$^3$INRNE,
 Bulgarian Academy of Sciences,
Tzarigradsko chauss\'ee 72, Sofia 1784, Bulgaria} 
\address{$^4$American University in Bulgaria,
 Sv. Buchvarova Str. 8, Blagoevgrad 2700, Bulgaria}
\address{$^5$African University of Science and Technology,
 P.M.B 681,
	Garki,
	Abuja F.C.T.,
	Nigeria.}
\ead{tdereli@ku.edu.tr, tpopov@aubg.edu}	
	

\begin{abstract}
We show that  the dynamical group of 
 an electron in a constant magnetic field is the group of symplectomorphisms 
$Sp(4,\mathbb{R})$. It is generated by the  
spinorial realization of
the conformal algebra $\mathfrak{so}(2,3)$ 
considered in Dirac's 
seminal paper "A Remarkable Representation of the 3 + 2 de Sitter Group". 
The symplectic group $Sp(4,\mathbb{R})$ is the double covering of
the conformal group $SO(2,3)$ of  2+1 dimensional Minkowski spacetime which  is in turn the dynamical group of a hydrogen atom in 2 space dimensions.  
The Newton-Hooke duality between the 2D hydrogen atom and 
the Landau problem  
is explained via the 
Tits-Kantor-Koecher  construction of the conformal
symmetries of the Jordan algebra
of  real symmetric $2 \times 2$ matrices.
The connection between the Landau problem
and the 3D hydrogen atom  is
elucidated by  the reduction of a Dirac spinor to a Majorana one in  the Kustaanheimo-Stiefel spinorial regularization. 
	
\end{abstract}

\section{Introduction}
{ In the heart of the quantum theory is the passionate history
	of the  hydrogen atom spectrum. By good fortune  the Coulomb potential describing the ``the action at a distance''  
	is the twin brother  of  the potential of the  Newton  universal interaction. Therefore the  quantum motion of electron in the hydrogen atom amounts to the quantization of the Kepler orbits. 
	In some poetical sense, the heavenly spheres mirror 
	the depth of the micro-cosmos.}

In his famous paper, Vladimir Fock \cite{Fock}
explained the  accidental degeneracies in the spectrum of
the hydrogen atom by the existence of an additional integral of motion given by  the Laplace-Runge-Lenz.
Fock's result allows  for a generalization. Namely,
a $n$-dimensional hydrogen atom whose Schr\"odinger equation for the bounded  states  in  momentum space is transformed by  stereographic projection  to the Laplace's equation on 
a $n$-dimensional sphere $S^n$\cite{Cordani}.

In this work we study the duality between the 2-dimensional (2D) hydrogen atom on one hand and
the Landau problem for  the quantization of the electronic orbits in the uniform
magnetic field   on the other:
\beq
\xymatrix
{
	\mbox{2D $e^-$  in electric field}
	& \stackrel{Newton-Hooke}{\leftrightarrow} & \mbox{2D $e^-$  in magnetic field
	} \ .
}
\label{NH}
\eeq
The  classical geodesic motion on the sphere, say on the  equator, is seen as the periodic circular Larmour motion 
in the magnetic field directed between the poles.
The Fock method applied for 2D hydrogen system with potential 
$-k/r$ sends orbital eigenfunctions to  spherical harmonics in momentum space
\cite{Cisneros,SW, BD, PP, TdC2} which are the harmonic functions of the sphere $S^2$, {\it i.e.,} eigenfunctions of 
an isotropic harmonic oscillator
with two degrees of freedom. The mapping 
from 2D harmonic oscillator to 2D Coulomb-Kepler problem
is known to be the Levi-Civita transform \cite{BD}. 
In this paper
we show that the  dynamical group of the Landau problem
is the double covering $Sp(4,\R)$ of the conformal group
$SO(2,3)$ of the 2D hydrogen atom and arises from the exponentiation of 
the remarkable  spinorial representation of $SO(2,3)$
found by Dirac  in his famous paper
"A Remarkable Representation of the 3 + 2 de Sitter Group"\cite{Dirac}.

Different dualities between Coulomb-Kepler motion and harmonic oscillators
has been already studied by
many autors\footnote{For comprehensive lecture notes and further references we send the reader to \cite{Ter}.}\cite{ BD, Cordani, Bars} in various dimensions,
however to the best of our knowledge the implications about
the Landau problem were not highlighted.
{A tremendous amount of work on the spectrum generating algebra of the hydrogen atom has been done by
	As{\i}m Barut and his collaborators \cite{BK,BB, BSW}(See also \cite{MM}).} Gradually it became clear that the  spectrum of 3D hydrogen atom 
carries a minimal (massless) representation with helicity 
$\lambda=0$ of the conformal group $SO(2,4)$. It allows for a dual description
in terms of 4D harmonic oscillator  via
a ladder $U(2,2)$-representation that has been put forward by Mack and Todorov\cite{MT}.
The correspondence with 
the massless $U(2,2)$-representation
having arbitrary helicity $\lambda$ (isomorphic  to a $SO(2,4)$-representation) turns out to be the quantization 
of the Kustaanheimo-Stiefel regularization \cite{KS} in celestial mechanics. This  correspondence holds true 
not only for the quantum Kepler motion
of electron in the field of the positively charged nucleus
but also for more general binary systems 
of coupled dyons \cite{BB,TdC3}.

A powerful approach to the conformal dynamical symmetries
of generalized quantum Kepler problems based on Jordan algebras
was proposed by Guowu Meng \cite{Meng}. 
To an Euclidean Jordan algebra one associates
a symmetric 
null ray cone
whose automorphisms form a conformal group.
From the minimal data of the Jordan algebra $\JJ^C_2$ of $2\times 2$ complex Hermitian matrices one is able to build  the whole conformal algebra $\mathfrak{so}(2,4)$ of 3D hydrogen atom\cite{Meng, Popov}.
We show that the reduction from 3D to 2D  hydrogen atom
dynamical group 
is done by simply imposing the reality condition on
the 
Jordan algebra of observables $\JJ^C_2$:
\beq
\label{pic}
\xymatrix{
	\mathfrak{co}(\JJ^\C_2) 
	& = & \mathfrak{so}(2,4)
	& \ar@{<-}^{Kustaanheimo-Stiefel}[r] 
	& &\mathfrak{su}(2,2) \\
	\mathfrak{co}(\JJ^\R_2)   \ar@{^{(}->}[u]^{x = x^t}
	& = & \mathfrak{so}(2,3)
	& \ar@{<-}^{Levi-Civita}[r] && \ar@{^{(}->}[u]^{\psi =\psi^c} \mathfrak{sp}(4,\R)}   \ .
\eeq
The projection
$\pi:\JJ^\C_2 \rightarrow \JJ^\R_2 $  simply projects out the complex Pauli matrix $\sigma_2$.
On the other hand the 3D hydrogen atom is 
in duality with  the 4D  harmonic oscillator (the quantization of  the harmonic motion on the sphere $S^3$)
the duality mapping  being the Kustaanheimo-Stiefel transformation \cite{KS}, the so called spinorial regularization removing collision Kepler orbits from
the phase space. Similarly the Levi-Civita transformation
connects the  2D harmonic oscillator
with the 2D hydrogen atom (\ref{NH}).
We found out that the spinorial reduction from 4D harmonic oscillator  (regularized motion on $S^3$)
to  2D harmonic oscillator (Landau problem, {\it i.e.,} 
motion on $S^2$)
is the reduction from Dirac to the Majorana spinor.
In other words on the spinorial side of the diagram
we also apply the reality condition $\psi=\psi^c$.
\section{Landau Problem and Harmonic oscillator}

An electron in an uniform magnetic field propagates 
on circular orbits  with Larmour frequency 
$\omega = \frac{e B}{mc}$. 
The Landau quantization of the circular orbits leads to an isotropic quantum harmonic oscillator with two modes.
The minimal coupling of the electron's charge density to the external magnetic field described by the 
electromagnetic vector potential $\bm {\Aa}$ leads to the Hamiltonian
\beq
\label{Ham}
H= \frac{1}{2m}\left( \bm{p} - \frac{e}{c}\bm{\Aa} \right)^2
=:\frac{1}{2m} \bm{P}^2 \ .
\eeq
The constant uniform magnetic field $ \bm  B= B \hat{z}$ along the $z$-direction
can be obtained from different potentials 
$\bm{\Aa}= (\Aa_x, \Aa_y)$
in the plane. We  choose  the symmetric gauge
\beq
\label{symg}
\bm \Aa = (\Aa_x, \Aa_y)= \frac{B}{2} (-y,x) \ ,  
\qquad \Aa_i=-\frac{B}{2}
\epsilon_{ij}x^j 
\eeq
but {most of} our conclusions are gauge independent. 
The gauge independent kinetical momenta $\bm P$
and the coordinates $\bm X$ of the center of the cyclotron motion are related to the phase space
 canonical coordinates  $(x,y,p_x,p_y)$ by 
\beq
\label{PP}
P_i = m\dot{x}_i
= m \frac{\partial H}{\partial p_i}=
 p_i - \frac{e}{c}\Aa_i, 
\qquad
X_i = x_i + \frac{1}{m \omega} \epsilon_{ij} P^j \ .
\eeq
The center of mass coordinates $\bm X= (X, Y)$ are  integrals of motion  $ {\dot{X}}=0= \dot{Y}$ and  decouple from the  system.  
One has two independent Heisenberg algebras
$[\bm P, \bm X]=0$
\[
[P_x, P_y]= i\frac{\hbar e}{c}B =i \hbar m \omega=\frac{i\hbar^2}{\ell^2}\ ,
\qquad  \qquad [X,Y]= -i \frac{\hbar}{m \omega} = -i \ell^2 \ . \]
Here $\ell$ stands for  the  magnetic length 
$\ell^2 = \frac{\hbar}{m \omega}=\frac{\hbar c}{eB}$.
We introduce also  dimensionless canonical coordinates of the phase space
\[
x = {{\sqrt{2}}} \xi \ell \qquad y= 
{{\sqrt{2}}} \eta \ell \qquad p_x = p_{\xi} \sqrt{m \omega \hbar }/ {{\sqrt{2}}} \qquad p_y = p_{\eta} \sqrt{m \omega \hbar }/  {{\sqrt{2}}}
\]
satisfying the canonical commutation relations 
\[
[\xi, p_\xi]=i = [\eta, p_\eta] \qquad [\xi, p_\eta]=0=[\eta,p_\xi] \ .
\]
The Hamiltonian $H$ in term of the new variables 
boils down to an isotropic harmonic oscillator Hamiltonian
plus a {"magnetic"} term proportional to the angular momentum:
\[
H= \frac{P^2_x+ P^2_y}{2m} = 
\frac{\hbar \omega}{{4}} \left\{ p_\xi^2+ p_\eta^2 + (\xi^2 +\eta^2)\right\} - \frac{\hbar \omega}{2}(\xi p_\eta-\eta p_\xi)
\ .
\]
The kinetic momenta $\bm P =\bm p -\frac{e}{c} \bm \Aa$
are quantized by
the energy creation and annihilation operators $a^{\pm}$.
The guiding center coordinates $X$ and $Y$
are integrals of motion, they are quantized by
the {\it magnetic translation operators}\footnote{The geometrical meaning of the Zak's magnetic translations in the Landau problem has been clarified in the work \cite{DP}.}
 $b^{\pm}$:
\beq
\label{osc}
\ba{lclcl}
a^{\pm} = \frac{a^\pm_x \mp i a^\pm_y}{\sqrt{2}}= 
\frac{-P_y \mp i P_x}{\sqrt{2 m \omega \hbar}}  \ ,&&
P_x=  \frac{ p_\xi +\eta}{{{\sqrt{2}}}}\sqrt{m\omega \hbar} \ ,
&&
P_y=    \frac{p_\eta -\xi}{{{\sqrt{2}}}} \sqrt{m\omega \hbar} \ ,
\\[4pt]
b^{\pm}=\frac{a^\pm_x \pm i a^\pm_y}{\sqrt{2}}= 
\frac{X \pm i Y }{\ell \sqrt{2}}  \ ,
&& X=  \left(\frac{\xi + p_\eta }{{{\sqrt{2}}}}  \right) \ell \ , &&
Y=  \left(\frac{\eta - p_\xi }{{{\sqrt{2}}}} \right) \ell
\ea
\eeq
where we have used two commuting Heisenberg algebras
with generators
\beq
a^{\pm}_x= ({\xi \mp i p_{\xi}})/{{\sqrt{2}}} \ , \qquad
a^{\pm}_y= ({\eta \mp i p_{\eta}})/{ {\sqrt{2}}}  \qquad
\mbox{such that}
\qquad
[a_i^-, a_i^+]=1 \ .
\label{axay}
\eeq



The operators $a^{\pm}$ shift between different energy levels of  the Hamiltonian 
\beq
H 
= \frac{\hbar \omega}{2} \{a^+, a^- \} \ ,
\qquad [{H}/{\hbar} \omega, a^{\pm}] = \pm  a^{\pm} \ , \qquad [H, b^\pm]=0 \ .
\eeq 
The ``zero mode" 
generators $b^\pm$ commute with $a^\pm$  
since the momenta $\bm P=(P_x, P_y)$ commute with
the guiding center coordinates $\bm X=(X,Y)$.
Hence the magnetic translations $b^\pm$  are responsible for the degeneracy of the Landau levels.
The angular momentum  operator $L_z=xp_y - yp_x$ is the generator of the rotational symmetry, the operators $b^\pm$
increase (decrease) the angular momentum eigenvalue  
\[
\frac{L_z}{\hbar} =
\frac{1}{2} \{b^-, b^+ \} - \frac{1}{2} \{a^-, a^+ \}\ ,
\qquad [\frac{L_z}{\hbar}, b^\pm] = \pm b^\pm \ , \qquad [L_z, H]=0 \ .
\]
The angular momentum  commutes with the energy operator $H$,
however, there is a larger accidental group $SO(3)$ of transformations
preserving the energy. Its origin is rooted 
in the analog of the Laplace-Runge-Lenz vector
in the Landau problem, the so called magnetic LRL vector 
\cite{LandauLRL} 
which is an integral of motion beside the angular momentum.  We proceed by exploring the dynamical group of the Landau problem   in a more  systematic way.
\section{Dirac dynamical  algebra
 $\mathfrak{so}(2,3)$
}

In his seminal paper 
"A Remarkable Representation of the 3 + 2 de Sitter Group"\cite{Dirac}  Dirac came out with a  realization of $\mathfrak{so}(2,3)$
which is quadratic in the phase space coordinates on a plane: $p_{\xi}, 
p_{\eta}, \xi, \eta$. The symplectic group $Sp(4,\R)$ is a natural
symmetry for the Landau problem since the magnetic field is encoded into the symplectic form of the  phase space.

We now show that  
the Dirac's algebra $\mathfrak{so}(2,3)$
of the De Sitter group $SO(2,3)$ can be thought of as the
group of symplectomorphisms, in view of the isomorphism
$$SO(2,3)\cong Sp(4,\R)/ \Z_2 $$ 
thus yielding a dynamical group
of the Landau's problem. The group $SO(2,3)$
has a homogeneous space
$dS_4$  defined by the Dirac quadric
	in 5-dimensional  flat ambient space with coordinates 
	$y^A$ 
\[
y_{-1}^2 +y_0^2 -y_1^2 - y_2^2 - y_3^2  = R^2 = 
\eta_{ab}y^a y^b \ ,
\qquad \eta_{ab} = diag(+1,+1,-1,-1,-1) \ .
\]
The group of motions of
the 4-dimensional hyperboloid  is the orthogonal de Sitter group  $SO(2,3)$, the conformal group of the
Minkowski space $\R^{1,2}$.
The  generators $m_{ab}$ of the Lie algebra  $\mathfrak{so}(2,3)$ 
satisfy the commutations relations 
\beq
\label{co}
[m_{ab}, m_{cd}] = 0 \ , \qquad \qquad
[m_{ab}, m_{bc}] = - i \eta_{bb} m_{ac} 
\eeq
where the indices $a$, $b$, $c$, $d$ are assumed to be all distinct from the set $\{-1,0,1,2,3 \}$. 

The Dirac's remarkable representation \cite{Dirac} is given  by  the following quadratic  generators\footnote{
We adopt a different convention for the indices of the matrix $m_{ab}$,
the mapping between our convention and the 
Dirac's one  reads $\{1,2,3,-1,0 \} \rightarrow \{1,2,3,4,5 \}$.}
 alternatively  written with the Heisenberg algebras
generators $a_i^\pm$ 
 (\ref{axay}):
\beq
\ba{lcccc}
m_{12} &=&
 \demi ( \xi p_{\eta}-\eta p_{\xi})
&=&\frac{1}{2i} \left(a_x^+ a_y^- -a^+_y a_x^-\right) \ ,\\[4pt]
m_{23} & = &\frac{1}{4}\left(p^{2}_{\xi}-p^{2}_{\eta}+\xi^2-\eta^2\right)
&=&-\demi\left(a^+_{x}a^-_{x}-a^+_{y}a^-_{y}\right) \ ,\\[4pt]
m_{31}&=&-\demi\left(\xi\eta+p_{\xi}p_{\eta}\right)
&=&-\demi\left(a^+_{x}a^-_{y}+a^+_{y}a^-_{x}\right) \ ,\\[4pt]
m_{1-1}&=&\demi\left(\xi\eta-p_{\xi}p_{\eta}\right)
&=&\demi\left(a^{-}_{x}a^{-}_{y}+a^{+}_{x}a^{+}_{y}\right) \ ,\\[4pt]
m_{2-1}&=&\frac{1}{4}\left(\xi^2-\eta^2+p^2_{\eta}-p^2_{\xi}\right)
&=&\frac{1}{4} \left(a^-_{x}a^-_{x}+a^{+}_{x}a^{+}_{x}- a^-_{y}a^-_{y}-a^+_{y}a^+_{y}\right)\\[4pt]
m_{3-1}&=&\demi\left(\xi p_{\xi}+ \eta p_{\eta}\right)-\frac{i}{2}
&=&\frac{i}{4}\left(a^+_{x}a^+_{x}-a^-_{x}a^-_{x}+a^+_{y}a^+_{y}-a^-_{y}a^-_{y}\right) \ ,\\[4pt]
 m_{01}&=& \frac{i}{2}\left(\xi p_{\eta}+\eta p_{\xi}\right)
&=&-\demi\left(a^+_{x}a^+_{y}-a^-_{x}a^{-}_{y}\right) \ ,\\[4pt]
m_{02}&=&\demi\left(\xi p_{\xi}-\eta p_{\eta}\right)
&=&\frac{i}{4}\left(a^+_{x}a^+_{x}-a^-_{x}a^-_{x}-a^+_{y}a^+_{y}+a^-_{y}a^-_{y}\right) \ ,\\[4pt]
m_{03}&=&
\frac{1}{4}\left(p^2_{\xi}+p^2_{\eta} -\xi^2-\eta^2 \right)
&=&\frac{1}{4}\left(
a^+_{x}a^+_{x}+a^-_{x}a^-_{x}+a^+_{y}a^+_{y}+a^-_{y}a^-_{y}\right) \ ,\\[4pt]
m_{-10}&=&\frac{1}{4}\left(p^2_{\xi}+p^2_{\eta}+\xi^2+\eta^2\right)&=&\demi\left(a^+_{x}a^-_{x}+a^-_{y}a^+_{y}\right) \ .\\[4pt]
\ea
\label{diracgen}
\eeq
Therefore the Dirac $\mathfrak{so}(2,3)$-representation is rooted in the oscillator algebra with two modes $a_x^{\pm}$
and $a_y^{\pm}$ \cite{BKN}.
We will further related these two modes $a_x^{\pm}$ and $a_y^{\pm}$ 
through eqs (\ref{osc}) to the energy and magnetic translation  creation and annihilation operators
$a^\pm$ and $b^\pm$.
 We will get the  Dirac
conformal algebra $\mathfrak{so}(2,3)$ playing the role of  infinitesimal  symplectomorphisms of the Landau problem.

It is convenient to introduce (anti)holomorphic coordinates $z$ ($\bar{z}$) on the phase space such that
\[
\ba{ccccc}
z &=& (x+ iy)/2\ell = {\frac{1}{\sqrt{2}}}( \xi +i \eta) &&
 \partial =\frac{\partial}{\partial z}= {\ell}\left(\frac{\partial}{\partial x}
-i \frac{\partial}{\partial y}\right) =
{\frac{1}{\sqrt{2}}}
\left(\partial_\xi - i \partial_\eta\right)  \ ,\\
\bar{z} &=& (x- iy)/2\ell = {\frac{1}{\sqrt{2}}}( \xi -i \eta)    &&
\bar{\partial}=
\frac{\partial }{\partial\bar{z}}= {\ell}\left(\frac{\partial}{\partial x}
+i \frac{\partial}{\partial y}\right) =
{\frac{1}{\sqrt{2}}}
\left(\partial_\xi + i \partial_\eta\right) \ ,
\ea
\]
where the momenta are related to the (anti)holomorphic derivatives $\partial$ ($\bar{\partial}$)  through
\[
\partial = \frac{i}{{{\sqrt{2}}}} (p_\xi - i p_\eta ) \ , \quad \bar{\partial} = \frac{i}{{{\sqrt{2}}}} (p_\xi + i p_\eta )  \ ,
\quad i p_\xi = {\frac{1}{\sqrt{2}}} (\partial +\bar{\partial}) \ , \quad
p_\eta={\frac{1}{\sqrt{2}}}(\partial - \bar{\partial}) \ .
\]
The energy and magnetic translation creation and annihilation
operators $a^{\pm}$ and $b^\pm$ in the Landau problem from equation (\ref{osc}) are then expressed in the holomorphic phase space coordinates
as follows
\footnote{The operators $a^{\pm}$ and $b^\pm$ yields
one more parametrization of the phase space $\R^4$.
The inverse transformation reads 
{ 
\[
\ba{ccccccc}
 {z} & = & {\frac{1}{\sqrt{2}}} (a^- + b^+ ) &&
 \bar{\partial} & = & {\frac{1}{\sqrt{2}}} 
(a^- - b^+ )\\
 {\bar{z}} & = & {\frac{1}{\sqrt{2}}}(a^+ + b^- ) &&
{\partial} & = &{\frac{1}{\sqrt{2}}} ( b^- - a^+  )
\ea
\]
}}
{
\beq
\ba{ccccccc}
a^- & =&\frac{1}{\sqrt{2}}({z} + \bar{ \partial})
&& b^- & =&\frac{1}{\sqrt{2}}({\bar{z}} +   \partial) \ ,\\
a^+ & =&\frac{1}{\sqrt{2}} ({\bar{z}} - \partial )
&& b^+ & =&\frac{1}{\sqrt{2}}({z} -  \bar{ \partial}) \ .
\ea
\eeq}
We are now able to recast the Dirac's generators $m_{ab}$ from eq (\ref{diracgen})  in the form
of quadratic polynomials of the creation and annihilation
operators $a^{\pm}$ and $b^\pm$ as follows
\beq
\ba{rcccc}
 m_{12} &=&
\demi (z \partial -\bar{z}\bar{\partial}) 
&=&  \frac{1}{4} \left(\{b^-, b^+ \}-  \{a^-, a^+ \}\right)\ ,\\[4pt]
m_{23} & = &
\frac{1}{4}(z^2+\bar{z}^2
-\partial^{2}-\bar{\partial}^{2})
&=&\frac{1}{4}\left(\{a^-,b^+\}+\{a^+,b^-\}\right) \ ,
\\[4pt]
m_{31}&=&
\frac{i}{4}(z^2-\bar{z}^2+\partial{^2}-\bar{\partial}^
{2})
&=&
\frac{i}{4}\left(\{a^-,b^+\}-\{a^+,b^-\}\right) \ ,\\[4pt]
m_{1-1}&=&
\frac{1}{4i}(z^2-\bar{z}^2 -\partial{^2}+\bar{\partial}^{2})
&=&\frac{i}{4}\left(a^{+}a^{+}-a^{-}a^{-} + b^{-}b^{-} - b^+b^+\right)\ ,\\[4pt]
m_{2-1}&=&
\frac{1}{4}(z^2+\bar{z}^2
+\partial^{2}+\bar{\partial}^{2})
&=& \frac{1}{4}\left(a^-a^- +a^+a^+ +b^-b^- +b^+b^+\right)
\ ,\\[4pt]
m_{3-1}
&=&-\frac{i}{2}\left(z\partial
+\bar{z}\bar{\partial}{-1} \right)&=&-\frac{i}{4}\left(\{a^-,b^-\}-\{a^+,b^+\}\right) \ ,\\[4pt]
m_{01}&=&
{{-}}
\demi\left(z\bar{\partial}- \bar{z}\partial\right)
&=&- {\frac{1}{4}}\left(a^-a^- + a^+a^+ -b^-b^- -b^+b^+\right) \ ,
\\[4pt]
m_{02}&=&-
\frac{i}{2}\left(\bar{z}\partial+z\bar{\partial}\right)
&=&-\frac{i}{4}\left(a^-a^- -a^+a^+ + b^-b^- -b^+b^+\right) \ ,\\[4pt]
m_{03}&=& \frac{1}{2} \left( z\bar{z}+\partial\bar{\partial}
\right) &=&
-\frac{1}{4}\left(\{a^+,b^+\}+\{a^-,b^-\}\right) \ ,
\\[4pt]
m_{-10}&=&
\frac{1}{2}\left( z\bar{z}-\partial\bar{
\partial} 
\right)
&=&\frac{1}{4}\left(\{a^-,a^+\}+\{b^-,b^+\}\right) \ .
\ea
\label{dynamo}
\eeq
{\bf Weyl Spinors and $Sp(4, \R)$.} The whole dynamic Dirac algebra $\mathfrak{so}(2,3) \cong
\mathfrak{sp}(4,\R)$
can be compactly represented  if we pack
the Landau's creation and annihilation  operators $a^{\pm}$ and $b^{\pm}$  into a two-component Weyl spinor 
\beq
\label{weyl}
\chi^\alpha = \left(\ba{c}b^- \\a^- \ea \right)=\frac{1}{\sqrt{2}} 
\left(\ba{c}
\bar{z} + {\partial}  \\  {z} +  \bar{\partial} \ea \right) \ , \qquad \chi^\ast_\alpha = (  b^+\, \, \, a^+) = 
\frac{1}{\sqrt{2}}(    {z} -  \bar{\partial} \, \, , \, \,  \bar{z}- {\partial} )
\ .
\eeq
Barut and Duru \cite{BD} have found that the conformal algebra
$\mathfrak{so}(2,3)$ generated by two oscillators (\ref{dynamo})
can be written by the spinorial operators 
\beq
\ba{rcccrcl}
m_{ij} &=& \demi \epsilon_{ijk} \chi^\ast \sigma_k^T \chi \ ,
 && m_{-1i} &=& \frac{i}{4} (\chi^\ast \sigma_i^T \epsilon^T (\chi^\ast)^T -
\chi ^T \epsilon \sigma_i^T \chi) \ ,
 \\[4pt]
m_{-10} &= & \demi ( \chi^\ast \chi +1) \ ,
&&  m_{0i} &=& \frac{1}{4}( \chi^\ast \sigma_i^T \epsilon^T (\chi^\ast)^T + 
\chi^T  \epsilon \sigma_i^T \chi) \ 
\ea 
\label{maj}
\eeq
with the help of the Pauli matrices $(\sigma_i)_{\alpha \beta}$ and the charge operator $\epsilon=i\sigma_2$. 
{ They have obtained the isomorphism of the spinorial
$\mathfrak{so}(2,3)$ representation (\ref{maj}) with  the spectrum generating algebra of the
 2D hydrogen atom. The latter spectrum has been derived as  a solution of infinite-component 
Majorana equation (see also the work of Stoyanov and Todorov \cite{ST}). }

The subalgebra generated by 
$m_{-10}$, $m_{-13}$ and $m_{03}$ is the radial subalgebra 
$\mathfrak{so}(1,2)$. The generator $m_{-10}$ is the conformal Hamiltonian attached  to the harmonic oscillator while the true hamiltonian is $\frac{H}{\hbar \omega}=m_{-10} - m_{12}$. 
As the motion of a harmonic oscillator is periodic 
the evolution  parameter $\tau$ attached to the operator
$m_{-10}$  will be compactified to $S^1$
and will be referred to as {\it conformal time $\tau$} .

The $\mathfrak{so}(3)$-algebra  spanned by the operators $m_{12}$, $m_{23}$,
 $m_{31}$ is commuting with 
 the conformal Hamiltonian $m_{-10}$
\[
[m_{-10}, m_{ij}]=0 \qquad i,j=1,2,3 \, \ . \]
The $\mathfrak{so}(3)$-algebra is   the  dynamical ``accidental" symmetry extending the rotational symmetry generated by the angular momentum operator $2 m_{12}=L_z/\hbar$. The rotational
group element  $R(\phi)=\exp i m_{12} \phi$ would actually live in a
spinor representation of $\mathfrak{so}(3)$.

\section{Jordan algebra toolbox}

The tight connection
between the  two  quantum problems sharing the same dynamical 
conformal symmetry $\mathfrak{so}(2,3)$: Landau problem 
and  2D hydrogen atom find their natural formulations 
in the setting 
of Jordan algebras.
We shall follow the ideas of Murat G\"unaydin \cite{G} to employ a Jordan algebra in the construction of a regular linear representation
of the conformal group. For the sake of completeness we first 
introduce the main notions related to
 Jordan algebra  in general and only then we specialize
to the important examples of the Jordan algebras
of the $2\times 2$ real Hermitian matrices $\JJ^\R_2$ 
and complex Hermitian matrices $\JJ^\C_2$ 
yielding the linear 
representation of 
 $SO(2,3)$ and  $SO(2,4)$, respectively.

{\bf Jordan algebras.}
A commutative  multiplication law 
$\circ: \JJ \times \JJ \rightarrow \JJ$, 
satisfying the Jordan identity
\begin{equation}
a\circ b=b\circ a \qquad 
a\circ \left(a^2\circ b\right)=a^2\circ \left( a\circ b\right)
\label{Jrd_2}
\end{equation}
defines a Jordan algebra $(\JJ, \circ)$.
Any associative matrix algebra (over $\R, \C, \H$) 
can be converted into a {\it special } Jordan algebra via the product 
$a\circ b=\frac{1}{2}\left( ab +b a\right)$,
 where $a b$ is the standard associative matrix multiplication. 

{\bf Jordan triple product and conformal group $Co(\JJ)$ representations. }
Any Jordan algebra $\JJ$ has a Jordan  triple product 
$(\bullet , \bullet , \bullet ):\JJ\times \JJ\times \JJ \rightarrow \JJ$ 
\beq
\label{JTP}
(abc)= a \circ (b \circ c) - b \circ (a \circ c)
+ (a\circ b) \circ c \ . 
\eeq

The Jordan triple product defines a Jordan triple system with identities 
\beqa
(abc)  &=&  (cba) \\
(ab(cdx)) - (cd(abx)) &=& (a (dcb)x) - ((cda)bx)
\label{5rel} \ .
\eeqa
For any pair of elements $(x,y)\in \JJ \times \JJ$ one has
a linear map $S_{x}^y: \JJ \rightarrow \JJ$  with a matrix fixed by the
 structure constants $(\Sigma_\mu^\nu)_\rho^\sigma$
of the Jordan triple product 
\beq
S_x^y(z)= (x y z) \qquad \qquad (e_\mu, e_\nu, e_\rho) = \Sigma_{\mu \rho}^{\nu \sigma}
 e_\sigma \ . 
\label{Jord_5}
\eeq

{ \bf Tits, Kantor and Koecher (TKK) construction of $\mathfrak{co} (\JJ)$. }
The conformal group $Co(\JJ)$ is generated by vector fields
closing a {\it  conformal }Lie algebra $\mathfrak{co} (\JJ)$. 
The general construction of a conformal Lie algebra $\mathfrak{co} (\JJ)$ from a given Jordan algebra $\JJ$ 
is due to Tits, Kantor and Koecher. 

Any Jordan triple system generated in $x\in \JJ$ 
 gives rise to a 3-graded Lie algebra $\mathfrak{co} (\JJ)$,
 endowed 
with an involution $\dagger$ via
$$(x,y,z):= [[x,y^\dagger],z] \ .$$
Conversely, any 3-graded Lie algebra
$[\mathfrak g_i, \mathfrak g_j]\subset  \mathfrak g_{i+j} $
(with $\mathfrak{g}_i=0$ when $ i\neq 0,\pm 1$)
endowed with a graded involution $\dagger$,
${\mathfrak g}_k^\dagger = \mathfrak g_{-k}$
determines
a Jordan triple system. 
The 3-graded Lie algebra $\mathfrak{co} (\JJ)$ 
has the graded decomposition
\beq
\mathfrak{co} (\JJ)  =\mathfrak g_{+1} \oplus \mathfrak{g}_0 
\oplus \mathfrak {g}_{-1} :=  
\JJ^\ast \oplus \mathfrak {str } (\JJ) \oplus { \JJ}
\eeq where the abelian
subalgebra  $\mathfrak g_{-1}$($\mathfrak g_{+1}$) is generated in the space $\JJ$($\JJ^\ast$).
The grading operator $D\in \mathfrak g_0 $ is  the 
dilatation $[D, g ]= k g $,  for any $g \in \mathfrak g_k $.
The grading alone implies that $\mathfrak g_{+1}$ and $\mathfrak g_{-1}$
are abelian Lie subalgebras  of $\mathfrak{co} (\JJ)$
and their mutual commutators belong to
$\mathfrak g_0$, $[\mathfrak g_{-1},\mathfrak g_{+1} ] \subset \mathfrak g_0$. 
It turns out that all elements in $\mathfrak g_0$ can be represented as commutators,  the structure algebra of $\JJ$,  
 $\mathfrak {str }(\JJ):=\mathfrak g_0 =[\mathfrak g_{-1},\mathfrak g_{+1} ] $
and  $\mathfrak g_{\pm 1}$ are its fundamental and antifundamental representations
in view of $$[\mathfrak g_0,\mathfrak g_{\pm 1} ]=\mathfrak g_{\pm 1} \ .$$
{\bf Coordinate-free definition of $\mathfrak{co} (\JJ)$.}
The relations\footnote{The  normalization $[U_a, U^b] = -2 S_a^b$ 
 is adopted for future convenience following \cite{Meng}. }
 of the conformal algebra $\mathfrak{co} (\JJ)$
are compactly written with the help of   the Jordan triple product 
conveying the essence of the TKK construction \cite{G} 
\beq
\ba{cccccclc}
[U_a, U^b] &=& -2 S_a^b \ ,&\qquad & [U_a, U_b] &=&0 \ ,& U_a\in \mathfrak g_{-1}\\[4pt]
 [S_a^b, U_c] &=& U_{(abc)} \ , &\qquad &  [S_a^b, S_c^d]&=& S_{(abc)}^d - S_c^{(bad)} \ ,& S_a^b \in  \mathfrak g_{0} \\[4pt]
[S_a^b, U^c]&=& -U^{(bac)} \ ,  &\qquad &
 [U^a, U^b] &= &0 \ ,&  U^b\in \mathfrak g_{+1}\ .
\ea
\label{tkk}
\eeq
The involution $\dagger$ acts as $U^\dagger_a = U^a$
and consequently  $(S_a^b)^\dagger = S_b^a$.

{\bf TKK for Jordan algebra $\JJ^\C_2=H_2(\C)$.}
The associative algebra of $2 \times 2$ Hermitian matrices with complex elements
$H_2(\C)$ has a basis of Pauli matrices $\sigma_\mu$.
The symmetric product $\circ$ gives rise to  the (special) Jordan algebra $\JJ^C_2$
\[
\sigma_i \circ \sigma_j:=\demi 
\{ \sigma_i, \sigma_j \}= \delta_{ij} \ ,
\qquad \sigma_0 \circ \sigma_i= \sigma_i \ ,
\qquad \sigma_0 \circ \sigma_0= \sigma_0 \ .
\]
An element $x \in H_2(\C) $ is parametrized by its 
Minkowski coordinates 
\[
x = x^\mu \sigma_\mu= x^0 1\!\!1+ x^i \sigma_i
\]
where $\sigma_0$ is the unit matrix $1\!\!1$ and $\sigma_i$ are the Pauli matrices. Hence
\beq
x = x^\mu \sigma_\mu=
\left( 
\ba{cc}
 x_0 + x_3 & x_1-i x_2 \\
 x_1 + i x_2 & x_0 - x_3
\ea
\right) \ , \quad \qquad x^\mu := \demi tr (x \sigma_\mu)
\label{sp}
\eeq
and the Minkowski metric $g$ with a signature $(+,-,-,-)$ is given by the determinant
$$
\det x = x_\mu x^\mu= (x^0)^2 -\delta_{ij} x^i x^j=
 g_{\mu \nu} x^\mu x^\nu  .$$

The evaluation of the Jordan triple structure constants 
$\Sigma_{\alpha \gamma}^{\beta \rho}$ in the $\JJ^\C_2$ basis 
\[
(\sigma_\alpha, \sigma_\beta, \sigma_\gamma) =
\Sigma_{\alpha \gamma}^{\beta \rho} \sigma_\rho=
\sigma_\alpha\circ (\sigma_\beta \circ \sigma_\gamma)-
\sigma_\beta\circ (\sigma_\gamma \circ \sigma_\alpha)
+ (\sigma_\alpha\circ \sigma_\beta) \circ \sigma_\gamma
\]
gives back  the concise formula,
\beq
\Sigma^{ \beta \rho}_{ \alpha \gamma} = 
\delta^\rho_\gamma \delta^\beta_\alpha  + \delta^\rho_\alpha \delta^\beta_\gamma - 
g^{ \beta \rho} g_{ \alpha \gamma} \ .
\label{JTcoef}
\eeq
The latter formula establishes via TKK construction a dichotomy between Minkowski's spacetime Jordan algebra $\JJ_2^\C$ and the 3-graded Lie algebra $\mathfrak{so}(2,4)$.

Indeed the  Jordan triple product induces via the TKK construction (Eq. (\ref{tkk}))  a $\mathfrak{co} (\JJ^\C_2)$-representation \cite{G, Palm}
by the following vector fields:
\[
\ba{c|rlc|ll|ll|c}
\mathfrak{co} (\JJ) & $operator$  & \in\mathfrak{co} (\JJ) & &$mapping $ & & 
x-$rep basis $ \mathfrak{so}(2,4)&  & $deg($x$) $   \\ \hline
\JJ & U_a=-ia^\mu P_\mu  &\in \mathfrak{g}_{-1} 
 && x\mapsto a &   &P_\nu =i \partial_\nu &  & 0 
 \\[4pt]
\mathfrak{str}(\JJ) & S_a^b=
i a^\nu b_\mu S^\mu_\nu &\in  \mathfrak{g}_{0} 
&& x\mapsto (a, b, x) &&
S_\nu^\mu=-i\Sigma^{\mu \beta}_{\nu \alpha} x^\alpha
\partial_\beta 
 &
 & 1  
  \\[4pt]
\JJ^\ast&U^b
=i b_\mu K^\mu  &\in \mathfrak{g}_{+1}
& & x\mapsto -(x, b , x) &   &K^\mu=
i
\Sigma^{\mu \beta}_{\nu \alpha} x^\nu x^\alpha
\partial_\beta&  & 2 
\ea
\]
The 15 generators of the conformal group
$Co( {\JJ^\C_2})=SO(2,4)$ are obtained 
after the evaluation of the Jordan structure constants  $\Sigma^{\mu \beta}_{\nu \alpha}$ (\ref{JTcoef})
\beq
\ba{lccclcl}
\mathfrak {g}_{-1}& &-i P_\nu & = & \partial_\nu & & $translations$ \\
\mathfrak {g}_{0}&&
i{M^\mu}_\nu & = & - x^\mu \partial_\nu + x_\nu \partial^\mu && $Lorentz transformations$  \\
\mathfrak {g}_{0}& &
i D & = & x^\mu \partial_\mu && $dilatation$ \\
\mathfrak {g}_{+1}&&
i K^\mu & = & - 2x^\mu x^\nu  \partial_\nu + x^\nu x_\nu \partial^\mu && $special conformal transformations$
\ea
\label{conf}
\eeq
yielding a minimal representation the Lie algebra $\mathfrak{so}(2,4)$. 
The Lorentz generators	${M^\mu}_\nu$ are the basis of 
	the reduced structure algebra $\mathfrak{str}_0(\JJ^\C_2)= \mathfrak{so}(1,3)$ (generating the Lorentz norm $\det x$ preserving group). The structure algebra $\mathfrak{str}(\JJ^\C_2)$ is generated by $S^a_b$. When we choose  coordinates as in the table above  $\mathfrak{str}(\JJ^\C_2)$ is the span of the generators
	$S^\mu_\nu=\frac{i}{2} [K^\mu, P_\nu]=
	 {M^\mu}_\nu - \delta^\mu_\nu D$.
	Then the TKK construction (\ref{tkk}) for $\JJ^\C_2$ yields
	the commutation relations of the conformal algebra
	$\mathfrak{so}(2,4)$
	of the Minkowski space $\R^{1,3}$ ($\mu, \nu, \lambda=0,1,2,3$)
	\beq
	\label{conff}
	\ba{cccccll}
	[K_\mu, P_\nu]&=&2 i (g_{\mu \nu}D - M_{\mu \nu}) \ , &&
	[D,P_\mu]= iP_\mu \ , && [D,K_\mu]= - iK_\mu \ , \\[4pt]
	[K_\lambda, M_{\mu \nu}] &=& i 
	(g_{\lambda \mu} K_{\nu}- g_{\lambda \nu} K_{\mu} ) \ ,&& [D,M_{\mu \nu}]=0\ , &&[P_\mu, P_\nu]=0=[K_\mu, K_\nu] \ ,\\[4pt]
	[P_\lambda, M_{\mu \nu}] &=& i 
	(g_{\lambda \mu} P_{\nu}- g_{\lambda \nu} P_{\mu} ) \ , &&
	[M_{\mu \alpha},M_{\beta \nu}]=0\ ,&& [M_{\mu \alpha},M_{\alpha \nu}]=ig_{\alpha \alpha}M_{\mu \nu} \ . 
	\ea
	\eeq
	In the last two relations the indices are assumed distinct.
	
		The conformal inversion $I(x^0, \bm x)=
 \left(\frac{x^0}{x^2}, -\frac{ \bm x}{x^2}\right)$ is an involution, $I^2=1\!\!1$. It anticommutes with the grading operator, $ID=-DI$ 
 and induces the involution $\dagger$ through
$$K^\mu = I P_\mu I=P_\mu^\dagger \ .$$
The conformal inversion changes dimensions; namely, the length  to inverse length.

{\bf TKK for Jordan algebra $\JJ^\R_2=H_2(\R)$.}
The real symmetric matrices are spanned by the 
subset $\{ \sigma_0, \sigma_1, \sigma_3 \}$
of real Pauli matrices with coordinates $\{ y_0, y_1, y_2 \}$
\beq
y = \sum_{\mu=0,1,3}x^\mu \sigma_\mu=
\left( 
\ba{cc}
 y_0 + y_2 & y_1
\\
 y_1 
 & y_0 - y_2
\ea
\right) \ , \quad y^T=y \, .
\eeq
We simply skip the Pauli matrix $\sigma_2$ 
(which is not symmetric) from the $3+1$ Minkowski spinor $x=x^\mu \sigma_\mu$. Renaming the components  $x^0=y^0$, $x^1=y^1$and $x^3=y^2$ we end up with a real spinorial representation of the  Minkowski spacetime $\R^{1,2}$ such that the deteminant 
yields the metric
$$
\det y = y_{\tilde{\mu}} y^{\tilde{\mu}}= (y^0)^2 -(y^1)^2- (y^2)^2=
 {g}_{\tilde{\mu} \tilde{\nu}} y^{\tilde{\mu}} y^{\tilde{\nu} } \qquad \tilde{\mu}, \tilde{\nu}=0,1,2 \ .$$ 

Tits-Kantor-Koecher construction
applied to  
 the Jordan algebra $\JJ^\C_2$ and  
$\JJ^\R_2$ yields,  respectively, the conformal algebra $\mathfrak{so}(2,4)$ of the 
Minkowski spacetime  $\R^{1,3}$, eq. (\ref{conff}):
\beq
\mathfrak{co} (\JJ_2^\C)=\mathfrak{so}(2,4) =
\underbrace{(\JJ_2^\C)^\ast}_{K^\mu} \oplus 
\overbrace{ (\underbrace{\mathfrak{so}(1,3)}_{M^\mu_\nu} \oplus \underbrace{{\R}}_{D})}^{
\mathfrak {str}(\JJ^\C_2) } \oplus  \underbrace{\JJ_2^\C}_{P_\nu} 
\label{so24}
\eeq
and the conformal algebra $\mathfrak{so}(2,3)$ of the Minkowski space $\R^{1,2}$, eq. (\ref{conff}):
\beq
\label{so23}
\mathfrak{co} (\JJ_2^\R)=\mathfrak{so}(2,3) =
\underbrace{(\JJ_2^\R)^\ast}_{K^{\tilde{\mu}}} \oplus 
\overbrace{ (\underbrace{\mathfrak{so}(1,2)}_{M^{\tilde{\mu}}_{\tilde{\nu}}} \oplus \underbrace{\R}_{D})}^{
\mathfrak {str}(\JJ^\R_2) } \oplus  \underbrace{\JJ_2^\R}_{P_{\tilde{\nu}}} \ . 
\eeq
{\bf Geometry of null cones.}
In his famous Erlangen program F\'elix Klein associated any geometric  space with its group of motion, {\it i.e.,} its underlying group of symmetries. 
\begin{prop} Let the compactified Minkowski  space
 ${\cal M}_{1,3}={{\cal N}}/\R^\ast$  be the space of  the isotropic rays in $\R^{2,4}$ (proportional isotropic vectors are identified)
\[
{\cal N} = \{ \vec{x}\in \R^{2,4}|   x^2_{-1} + x^2_{0}- x^2_1 - x^2_2 - x^2_3 - x^2_5
=0; x\neq 0 \} . 
\]
The compactified null cone 
${\cal M}_{1,3}={{\cal N}}/\R^\ast \cong (S^1 \times S^3)/\Z_2
$
 is a homogeneous space for the conformal group $SO(2,4)$
generated in the Lie algebra $\mathfrak{co} (\JJ_2^\C)=\mathfrak{ so}(2,4)$, eqs (\ref{conf}). 

The intersection of ${\cal M}_{1,3}$ by hyperplane $x_2=0$ yields  the space 
${\cal M}_{1,2}={\tilde{\cal N}}/\R^\ast$ of the isotropic rays in $\R^{2,3}$ 
\[
\tilde{\cal N} = \{ \vec{y}\in \R^{2,3}|   y^2_{-1} + y^2_{0}- y^2_1 - y^2_2 - y^2_3 
=0; y\neq 0 \} 
\]
where 
the remaining coordinates after the reduction
are renamed according to
$$\{x_{-1}, x_0, x_1, x_3,x_5\} \leftrightarrow
\{y_{-1}, y_0, y_1, y_2,y_3 \} \ . $$
The compactified null cone ${\cal M}_{1,2}={\tilde{\cal N}}/\R^\ast
\cong (S^1 \times S^2)/\Z_2$ of the isotropic rays in $\R^{2,3}$
is a homogeneous space for the conformal group $SO(2,3)$ generated in 
the Lie algebra
$\mathfrak{co} (\JJ_2^\R)=\mathfrak{ so}(2,3)$
(see eqs (\ref{conf})) . 
\end{prop}
In this sense,
the geometry of the compactified Minkowski
spaces ${\cal M}_{1,2}=(S^1\times S^2)/{\mathbb Z}_2$ and 
${\cal M}_{1,3}=(S^1\times S^3)/{\mathbb Z}_2$ are associated with the conformal symmetry algebras 
$\mathfrak{so}(2,3)$ and $\mathfrak{so}(2,4)$
of the Jordan algebras of real and complex hermitian $2 \times 2$ matrices.
The intersection with the plane $x_2=0$ corresponds to the reduction of the  complex Jordan algebra  to real $\JJ^\R_2$
Jordan algebra $\JJ^\C_2$
\[
\JJ^\R_2 = \{ y \in \JJ^\C_2 | y^T= y \} \ , \ 
\]
thus reducing the conformal symmetry from $\mathfrak{so}(2,4)$ to $\mathfrak{so}(2,3)$.

The advantage of  the TKK construction
of the Jordan algebra symmetries is that it yields explicitly a conformal group representation
 of the conformal  spacetime symmetries as linear transformations 
of a null-ray cone. 


\section{Hydrogen atom from 3D to 2D}

The non-relativistic hydrogen atom in three space dimensions is central in the development of the  quantum mechanics.   
 Barut and collaborators 
(see e.g. \cite{BB, BSW}) have shown that  the states in the  3D hydrogen atom
 spectrum transform in a helicity zero massless irreducible representation \cite{MT} of the  dynamical group $SO(2,4)$
with generators
\beq
\ba{rclcrclcrcl}
 \bm L &=& \bm r \times \bm p &\qquad &B_0- A_0 & = & r &\qquad & B_0+ A_0 & = & r{\bm p}^2
\\
\bm \Gamma & =& r \bm p& & \bm B - \bm A  &=& \bm r
 & &
\bm B + \bm A  &=& \bm r {\bm p}^2 -
2\bm p (\bm r \cdot \bm p) \\ 
D &=& \bm r\cdot \bm p - i  &&&&
\ea \ .
\label{dyn}
\eeq
 On the other hand the conformal symmetry  $SO(2,4)$ of the hydrogen spectrum is also the group of
causal space-time automorphisms of the Minkowski space $\R^{1,3}$.  The dictionary between the two
is given by the table
\[
\begin{tabular}{|c|c|c|}
\hline
 $\mathfrak{co}(\JJ)$& space-time cone & 
 hydrogen atom \\
\hline
$\mathfrak{g}_0$ & $M_{\mu \nu}$ ; $D$ & $L_{i j}$, $\Gamma_i$; $D$\\
$\mathfrak{g}_{-1}$ & $K_\mu$ &$B_{\mu} +A_{\mu}$\\
$\mathfrak{g}_{+1}$ & $P_\mu$ & $B_{\mu} -A_{\mu}$\\
\hline
\end{tabular} \ .
\]

The 
dynamical group $SO(1,2)$ generated by
$\{ A_0, D, B_0 \}$
is the  ``radial'' group of the conformal transformations of the time coordinate $\R^{0,1}$ \cite{Jack}
\[
\ba{cccclclc}
t &\rightarrow &t^\prime&=&t+t_0  && P_0 = B_0- A_0 &
$time translations$\\
t &\rightarrow &t^\prime&=&\lambda t &&D & $time dilation$\\
t &\rightarrow &\frac{1}{t^\prime}&= & 
\frac{1}{t} +\frac{1}{t_0} && K_0= A_0 + B_0&
$special conformal$
\ea \ .
\]
The conformal transformation group $SO(1,2)$ arises from the 
TKK construction of the  Jordan algebra  $\JJ^\R_1 =\R$
yielding
 $$\mathfrak{co}(\R)=\mathfrak{so}(1,2)=
(\JJ_1^\R)^\ast \oplus 
\mathfrak {str}(\JJ^\R_1)  \oplus 
\JJ_1^\R= \R K_0 \oplus \R D \oplus \R P_0 \ .
$$
The sign of  energy chooses different
$SO(1,2)$-generator to be a conformal hamiltonian:
\[
\ba{rcccccl}
B_0 &=& r(p^2 + 1)/2 &\qquad \quad& E<0 && $bound states$\ , \\
A_0 &=& r(p^2 - 1)/2 && E>0 && $scattering states$ \ ,\\
A_0+B_0 &=& r p^2  && E=0 && $free motion$ \ .
\ea
\]
The conformal $SO(1,2)$ group can be seen also as the even generators of the Schr\"odinger group in the context of non-relativistic conformal symmetries (for the correspondence 
between Schr\"odinger and conformal group  see \cite{Henkel}).

The 15 generators of the 3D hydrogen dynamical group $SO(2,4)$
 act by linear transformations  of the null-ray cone. They
have been identified in \cite{Popov} (see the table above)
 with the causal automorphisms of the light cone $\mathfrak{co}(\JJ^\C_2)$, see eqns  (\ref{so24}, \ref{conf}) 
\beq
\label{bar}
\left(
\ba{cccccc}
0& B_0 
 & B_1  & B_2  & B_3  & D 
\\
&0& \, \Gamma_1 \, &  \Gamma_2  & \Gamma_3  & A_0 
 \\
& &0& \, L_3 \,  &\, -L_2 \, & \,A_1\,  \\
& & &  0  & L_1  & A_2  \\
& & &     &   0  & A_3   \\
& & &     &      &  0          
\ea
\right)=
\left(
\ba{cccccc}
0& \, L_{-10}\, &  L_{-11}&  L_{-12} &   L_{-13}
&  L_{-15} \\
&0& \, L_{01} \,  &\, L_{02} \, & \,L_{03}\, &\, L_{05}\, \\
& &0& \, L_{12} \,  &\, L_{13} \, & \, L_{15}\,  \\
& & &  0  & L_{23}  & L_{25}  \\
&& &     &   0  & L_{35}  \\
&& &     &      &  0       \\
\ea
\right) \ .
\eeq
Stated differently  the compactified Minkowski space 
${\cal M}_{1,3}=(S^1\times S^3)/{\mathbb Z}_2$ carries
a minimal representation of the conformal group $SO(2,4)$
stemming from the TKK construction of the Jordan algebra
$\JJ^\C_2$.

The 
 conformal algebra   generators
$L_{AB}$
satisfy the commutation
 relations 
\beq
[L_{AB},L_{CD}]=- i( \eta_{AC}L_{BD}+ \eta_{BD}L_{AC}
-\eta_{AD}L_{BC} -\eta_{BC}L_{AD})
\label{conf+}
\eeq
where  the set of indices 
contains the auxiliary indices $-1$ and $5$ in addition to  the spacetime indices $\mu=0,1,2,3$
$$L_{AB} \in \mathfrak{so}(2,4) \ ,\qquad \eta_{AB}=diag(1,1,-1,-1,-1,-1) \ ,\qquad  A, B \in \{-1, 0, 1,2,3, 5 \}  \ .
 $$ 
While reducing to the  spacetime ${\mathbb R}^{1,2}$ with indices
$\mu=0,1,2$ 
we get an algebra representation equivalent  to the Dirac's remarkable $\mathfrak{so}(2,3)$-representation (\ref{co})
$$L_{ab} \in \mathfrak{so}(2,3) \ ,\qquad \eta_{ab}=
diag(1,1,-1,-1,-1)
\ ,\qquad  a,b \in \{-1, 0, 1,2,3=5 \}
\footnote{The auxiliary index $5$ comes with the Dirac matrix $\gamma_5=\gamma_0 \gamma_1\gamma_2\gamma_3$. In $\R^{1,2}$
we have  $\gamma_5=\gamma_0 \gamma_1\gamma_2$ 
and similarly $i \sigma_3=\sigma_0 \sigma_1 \sigma_2$ so we adopt  a double notation $3=5$.
}  \ .
 $$

When the motion of the electron is constrained  to a plane we obtain a system which we will refer to as 
2D hydrogen atom, the reduction of the usual 3D atom to 
two space dimensions. The
dynamical algebras of the 2D hydrogen atom and the Landau problem   are isomorphic, the isomorphism is  simply the transposition
\beq
L_{ab} \quad \leftrightarrow \qquad m_{ba}=-m_{ab} \ .
\eeq

The isotropic rays in the five dimensional space $\R^{2,3}$
carry a linear representation of the conformal group $SO(2,3)$
with generators  :
\[
L_{ab} =
\left(
\ba{ccccc}
0& B_0 & B_1  & B_2  & D \\
&0& \, \Gamma_1 \, &  \Gamma_2  &   A_0
 \\
& &0& \, L_3 \,   \, & \,A_1\,  \\
& & &  0   & A_2  \\
& & &      &  0          
\ea
\right)
=
\left(
\ba{ccccc}
0& \, L_{-10}\, &  L_{-11}&  L_{-12} 
&  L_{-15} \\
&0& \, L_{01} \,  &\, L_{02} \,  &\, L_{05}\, \\
& &0& \, L_{12} \,  & \, L_{15}\,  \\
& & &  0  & L_{25}  \\

&& &           &  0       \\
\ea
\right) .
\]
The dynamical symmetry group of the $2D$ hydrogen atom is 
the 
 conformal group 
$SO(2,3)$ stemming from the TKK algebra $\mathfrak{co}(\JJ^\R_2)$ (\ref{so23}). The compactified Minkowski space  ${\cal M}_{1,2}
\cong (S^1 \times S^2)/\Z_2$ is a homogeneous space
for the dynamical
$SO(2,3)$ group of the 2D hydrogen atom. We conclude that 
in  the Jordan algebra language the reduction
from $SO(2,4)$ to $SO(2,3)$ is projecting
the complex Jordan algebras $\JJ^\C_2$ to the real $\JJ^\R_2$.

{ The sphere $S^3$ in ${\cal M}_{1,3}$  arises as compactification of the flat momentum space via the Cayley transform whereas $S^1$ stays for the compatified time coordinate. 
The maximally compact subgroup $SO(4)\subset SO(2,4)$ stabilizes the  sphere $S^3$ interpolating between  states with
equal energy. A dual point of view is advocated in a recent paper \cite{KPV}  where $S^3$ in ${\cal M}_{1,3}$ is thought as a configuration space of the quark-antiquark system. 
In that way the compactified Minkowski space ${\cal M}_{1,3}$ becomes a toy model for the  simplest QCD system: a meson. In other words one obtains  the  "QCD hydrogen atom" where the interaction potential is the curved analogue of the Coulomb potential, inherently introduced by  the Green function of the Laplace-Beltrami operator on $S^3$. On closed surfaces charges
appear in pairs thus the charge neutrality is naturally leading to a confinement. The phenomenologically observed degeneracies of the spectrum of meson masses can be then attributed to  the conformal symmetries \cite{KC} of the compactified Minkowski space $ (S^1 \times S^3)/\Z_2$
seen as a configuration space for one color charge degree of freedom. 
From that perspective the mesons are the QCD cousins of the
system "electron-constant magnetic field" living in the space ${\cal M}_{1,2}\cong (S^1 \times S^2)/\Z_2$.}

\section{ Reduction of Kustaanheimo-Steifel transform 
}

The Kustaanheimo-Stiefel transform \cite{KS} in celestial mechanics removes  the singular trajectories due  to  binary  collisions from the phase space of the 3D Kepler motion and
estabilishes a correspondence between the 4D isotropic harmonic oscillator 3D Coulomb-Kepler problem. 
 The spectrum of bounded states for the 3D hydrogen atom 
({\it i.e.} the quantum Kepler problem) is then 
symplectically  equivalent 
to the spectrum of the  harmonic oscillator 
with 4 bosonic modes. One has an inclusion $SU(2,2)\subset Sp(8,\R)$ \cite{MT}.

The elliptic Kepler orbits for negative energies $E<0$ correspond to  geodesic motion on  3-dimensional sphere $S^3$.
The sphere $S^3$ arises
through the stereographic projection of the momenta. A  great circle in $S^3$
is the hodograph\footnote{The hodograph is a curve drawn by the
velocity vector, that is, the trajectory in the momentum space. It turns out the the hodographs of Kepler orbits for hyperbolic ($E>0$) and parabolic $E=0$ motions are  
segments of circles.} of an elliptic orbit \cite{Milnor}. The regularized Kepler orbits live on   
  the cotangent bundle $T^\ast S^3$ to the sphere $S^3$ with the zero section removed \cite{Cordani} 
	$$T^+S^3 = T^\ast S^3 -\{ 0_{sec} \}  \ . $$

\begin{deff}
The Kustaanheimo-Stiefel ($KS$) transform  is the mapping
between the cotangent bundles 
(with north poles deleted) \footnote{
We adopt the notation $\R^\ast=\R - \{ 0 \}$. }
$$KS: \quad T^+ S^3 \rightarrow T^+ S^2 \qquad
\subset
(\R^\ast)^4 \times \R^4  \rightarrow
(\R^\ast)^3 \times \R^3$$
where the 4D harmonic oscillator phase space 
coordinates $(u, w)\in (\R^\ast)^4 \times \R^4$ are related with  the 3D Kepler phase space $(x,p) \in (\R^\ast)^3 \times \R^3$ coordinates
through the Hopf fibration map (see eq. (\ref{Hopf}))
\beq
\ba{ccc}
x_1 & = & u_1 u_3 + u_2 u_4  \ ,
\\
x_2 & = & u_2 u_3 -u_1 u_4  \ ,  \\
x_3 & = &- u_1^2 - u_2^2 + u_3^2 + u_4^2 \ ,
\ea 
\eeq
extended with the derived momentum relations ($|\bm z|^2=u_1^2+u_2^2+u_3^2+u_4^2$)
\beq
\ba{ccc}
p_1 & = & -(u_1 w_3+ w_1 u_3 + u_2 w_4 + w_2 u_4)/|\bm z|^2 \ ,
\\
p_2 & = & -(u_2 w_3 + w_2 u_3-w_1 u_4 -u_1 w_4)/|\bm z|^2  \ ,  \\
p_3 & = & (u_1 w_1 + u_2w_2 - u_3w_3 - u_4w_4)/|\bm z|^2 \ .
\ea \ .
\label{KS}
\eeq
The coordinates on $T^+S^3$ are subject to the constraint
\[
K=u_1 w_2 - u_2 w_1 + u_3w_4 - u_4 w_3=0 \ .
\]
\end{deff}
This constraint  is in fact a constant of motion as a sum
of two angular momentum components. The non-vanishing of
the constraint $K$ is playing a crucial role in the generalizations of the magnetised problem of a electric charge
in the field of a Dirac monopole and systems of two dyonic particles \cite{TdC3}.

The Hopf  fibration is essentially  representing a 3D vector $\bm x\in \R^3$ as a ``square root'' of a spinor $\bm z\in \C^2$ in view of
\beq
|\bm x|= |\bm z|^2 \ , \qquad |\bm x|=\sqrt{x_1^2+x_2^2+x_3^2}
\ , \qquad \bm z = \left(\ba{c} u_1+ iu_2 \\ u_3 + i u_4\ea \right) \ ,\qquad  |\bm z|^2=\bm z^\dagger \bm z .
\label{Hopf}
\eeq
The KS transformation can be seen as  a phase space extension of the Hopf fibration
\[
0\rightarrow S^1 \hookrightarrow S^3  \rightarrow S^2 \rightarrow 0 \ ,
\]
the kernel consists of  the spinors $e^{i\theta}\bm z \in S^1$.

Reduction of KS transform  to the Levi-Civita transform  is done by the choice
\[
u_2=u_4=0 \ , \qquad w_2=w_4=0
\]
which amounts to taking a square root of a vector in the 2D plane  $x_1=\eta$ and $x_3=\xi$.

The Levi-Civita transform stems from the change to 
parabolic coorindates  $u_1$ and $u_3$ in the complex plane
%
\beq
\xi+i\eta=Z^2 \ ,\qquad 
\ba{ccccc}
\xi &=& u_1^2 -u_3^2 \ ,&\qquad&
 p_\xi  =  (u_1 w_1- w_3 u_3)/|Z|^2 \ ,
\\
\eta &=& 2u_1 u_3 \ ,&& p_\eta  =  -(u_1 w_3+ w_1 u_3)/|Z|^2
\ea
\label{LC}
\eeq
where the real spinor $\psi =
\left( \ba{c}u_1 \\ u_3 \ea \right)
$ is written with one complex number $Z=u_1+iu_3$. 
Levi-Civita mapping is then a sympletic extension of the 
the trivial Hopf fibration
\[
0\rightarrow S^0 \hookrightarrow S^1  \rightarrow S^1 \rightarrow 0 \ 
\]
where the dimension zero sphere $S^0=\Z_2$ is in the kernel
thus reflecting the fact that any pair of parabolic (spinor)  coordinates $(u_1,u_3)$ and $(-u_1,-u_3)$ parametrize one 
and the same point $(\xi, \eta)$ in the complex plane.

The Levi-Civita transform (\ref{LC}) is then a 2-to-1 
mapping between  2D harmonic oscillator phase-space $(u_1,u_3,w_1,w_3)$ and the phase space of the 2D Kepler problem 
$(\xi, \eta, p_\xi, p_\eta)$. 
Upon quantization it yields the Newton-Hooke duality 
between the Landau problem and the 2D hydrogen atom (\ref{NH}).
We are now going to show that the phase space $(u_1,u_3,w_1,w_3)$ is naturally parametrized by a 
 Majorana spinor in dimension 4.

\section{Ladder   $U(2,2)$ representation}

Let $\psi= (\psi^\alpha)$ be operator-valued Dirac spinor with 4 components satisfying the canonical commutation relations
\[
[\psi^\alpha, \bar{\psi}_\beta] = \delta^\alpha_\beta \ ,
\qquad \qquad  [\psi^\alpha, {\psi}^\beta] = 0 \ . 
\] 
A canonical representation of a pair of 2D harmonic oscillators with  two complex  variables $z^{\alpha}$ and the holomorphic derivatives $iw_\alpha=\frac{\partial \, \,}{\partial z^{\alpha}}$ is given by
\beq
\label{di}
\bm z= \left( \ba{r} z^1 \\ z^2\ea \right) \qquad 
\bm \partial=  \left( \ba{c} \frac{\partial\,\, }{\partial z^{1}} \\[4pt]
 \frac{\partial\,\, }{\partial z^{2}} \ea \right) \qquad
\psi = \left( \ba{r}  \bar{z} \\ {\partial} \ea \right) \qquad
\bar{\psi} = 
 (- \bar{ \partial } , z)
\ .
\eeq
The ladder representation of $U(2,2)$ \cite{MT} is a 
spinorial representation realized by 
the operators\footnote{Here the summation on spinorial indices is implicit.}
\beq
J^{AB}= \bar{\psi} \sigma^{AB}\psi  \ ,\qquad \mbox \quad
C_1=\bar{\psi} \psi 
\label{dir}
\eeq 
where the $4\times 4$ matrices $\sigma^{AB}$ 
close a defining representation of $\mathfrak{su}(2,2)$:
these are the matrices in $\mathfrak{sl}(4, \C)$
preserving a pseudo-Hermitian form $\beta$ with signature $(++--)$
\[
(\phi,\psi) = \phi_1^\ast \psi_1 +\phi_2^\ast \psi_2-
\phi_3^\ast \psi_3-\phi_4^\ast \psi_4 = \phi^\dagger \beta \psi= \bar{\phi} \psi \ .
\]
The Hermitian matrix $\beta=\beta^\dagger$ depends on the basis, it fixes the choice of the Dirac matrix $\gamma_0:=\beta$ and  the invariance of the form implies
\[
\beta\sigma^{AB} \beta^{-1}= (\sigma^{AB})^\dagger  \ .
\]
The linear Casimir operator $C_1$ is  the center of $U(2,2)$.
It is represented by an integer multiple of the unit 
$ 1\!\! 1$:
$$C_1+2= -2\lambda =
z^\alpha \frac{\partial}{\partial z^{\alpha}} 
- \bar{z}^\alpha \frac{\partial}{\partial \bar{ z}^{\alpha}} 
 \qquad \qquad \lambda =0, \pm \demi, \pm 1, \ldots $$ 
where the half-integer  helicities $\lambda$ are  labelling
the zero-mass representations of $U(2,2)$.
The hydrogen atom is described by helicity $\lambda=0$ representation \cite{MT}.

A finite-dimensional representation of $\mathfrak{su}(2,2)$ 
is generated by the $4\times 4$ Dirac gamma matrices  
$\gamma^\mu$, 
\[
\{ \gamma^\mu,  \gamma^\nu \} = 2 \eta^{\mu \nu} \qquad
\mu, \nu=0,1,2,3 \ .
\]
The 15 $\mathfrak{su}(2,2)$-generators $\sigma^{AB} $ can be concisely written\footnote{
We get another concise expression  
$
\sigma^{AB} = \frac{i}{2} \gamma^A \gamma^B$ for $A<B$ in 
$ \{-1,0,1,2,3, 5 \}$ if we set the unit matrix multiplier $\gamma^{-1}:= i1\!\! 1$.}(see e.g. \cite{MT}) :
\beq
\ba{cclcccl} 
\sigma^{\mu \nu} &=& \frac{i}{4} [\gamma^\mu, \gamma^\nu]\ , &\qquad \qquad&\sigma^{-1 5}&=&- \demi \gamma^5 \ , \\[4pt]
\sigma^{\mu 5} &= & \frac{i}{4} [\gamma^\mu, \gamma^5]\ , &&
\sigma^{-1 \mu}&=&- \demi \gamma^\mu \ .
\ea
\eeq
Here we denoted  $\gamma^5:=\gamma^0\gamma^1 \gamma^2 \gamma^3$.
In view of the isomorphism $\mathfrak{su}(2,2)\cong \mathfrak{so}(2,4)$ the operators $\sigma^{AB}$
satisfy the $\mathfrak{so}(2,4)$ commutation relations (\ref{conf+}). They   give rise to a spinorial representation
of the conformal algebra $\mathfrak{so}(2,4)$, eq. (\ref{bar})
through  the Kustaanheimo-Stiefel correspondence 
between the quantum Coulomb-Kepler problem and the 4D harmonic oscillator
$$L_{AB}\qquad \leftrightarrow \qquad  J^{AB}=\bar{\psi} \sigma^{AB} \psi \ ,\qquad \qquad A,B\in \{-1,0,1,2,3, 5 \}  \ .$$

{
\begin{prop} 
The Majorana condition $\psi=\psi^c$ is reducing the 
Dirac $SU(2,2)$-spinor $\psi_\alpha$ to a real $Sp(4,\R)$-spinor. The  reality condition is essentially  reducing the 4D harmonic oscillator (Dirac spinor) to the 2D oscillator (Weyl spinor).
  \end{prop}
	
	{\bf Proof.} The Majorana condition is the invariance under
	the charge-conjugation involution $C=i \gamma^0 \gamma^2$.
We choose the Dirac  representation for the Clifford algebra generators $\gamma^\mu$
\[
\gamma^0 =\left(
\ba{cc} \sigma_0 & 0 \\
0 & - \sigma_0 
\ea \right) \ , \qquad
\gamma^i =\left(
\ba{cc} 0 & \sigma_i  \\
 - \sigma_i & 0
\ea \right) \ , \qquad
 C= i \gamma^0 \gamma^2= \left( \ba{cc} 0 & \epsilon \\
 \epsilon & 0
\ea \right) \ .
\]

A Majorana spinor is a Weyl spinor written in 4-dimensional form. The Majorana condition $\psi=\psi^c$ yields
\beq
\label{major}
\psi = 
\left(
\ba{c}
\chi \\
\epsilon^T (\chi^\ast)^T
 \ea \right) \ ,
\qquad 
\bar{\psi} = \psi^\ast \gamma^0=
\left(
\chi^\ast \quad
- \chi^T \epsilon
  \right) \ .
\eeq
The Dirac's remarkable spinorial $\mathfrak{so}(2,3)$-algebra represention (\ref{maj}) is nothing else than 
the ladder representation for the Majorana spinor $\psi=\psi^c$
\[
m_{ba} \quad \leftrightarrow \quad  J^{ab}= \bar{\psi}^c \sigma^{ab} \,\psi^c \qquad \qquad a,b\in \{-1,0,1,2,3 \} \ .
\]
 The  Weyl spinor  $\chi^{(\ast)}_\alpha$ has components  identified in eq. (\ref{weyl}) with
the Heisenberg algebra generators
 $a^\pm$ and $b^\pm$ in Landau problem.
Hence the dynamical algebra $\mathfrak{so}(2,3)$  of the Landau problem (\ref{dynamo}) is the span of
 the infinitesimal symplectomorphisms $\mathfrak{sp}(4,\R)$ of the phase space $(\xi,\eta,p_\xi,p_\eta)\in \R^4$ obtained
from the Majorana reduction 
of the ladder representation (\ref{dir}) of 
 $\mathfrak{su}(2,2) \cong\mathfrak{sp}(8,\R)\cap \mathfrak{so}(4,4)$. 

}

\section{Conclusion and outlook}

In celestial  mechanics the regularization is the transformation of the singular equations of the  Kepler motion
to the regular equations of the harmonic oscillator. 
The duality between the Newton's universal law and Hooke's
 law of harmonic motion has been know since the time of Newton
and Hooke \cite{Chandrasekhar}. Its reincarnation in quantum mechanics is
the duality between the hydrogen atom and the harmonic oscillator via the Fock's method of quantization.
At the quantum level, the Levi-Civita regularization induces the duality transform
between the 2D hydrogen atom and 2D harmonic oscillator,
whereas  the Kustaanheimo-Stiefel regularization  bridges
between  the 3D hydrogen atom and the 4D harmonic oscillator.
We have given an unified picture (\ref{pic}) of the dynamical groups in duality derived  from the conformal symmetries on Jordan algebras and the TKK construction. 
We point out that the dynamical group $Sp(4,\R)$ of
the quantum motion of an electron in a constant magnetic field
(Landau problem) is  mapped via the Levi-Civita regularization
to the conformal $SO(2,3)$ spectrum generating algebra of the
2D hydrogen atom. The symplectic group $Sp(4,\R)$ is a natural
symmetry for the Landau problem since the magnetic field is encoded into the symplectic form on the  phase space of the planar motion.
On the other hand the carrier of the electromagnetic interaction is the zero-mass photon propagating on the light cone in Minkowski spacetime $\R^{1,2}$ whose   causal automorphisms form the group  $SO(2,3)$. In a similar fashion,
the Kustaanheimo-Stiefel regularization connects
the zero-mass  spinorial representation of $SU(2,2)$ with 
the minimal $SO(2,4)$-representation of  light cone
automorphisms in Minkowski spacetime $\R^{1,3}$.

The merit of the Jordan
algebra formalism is that the reduction from 
3D to 2D hydrogen atom is done by the reduction
of the Jordan algebra $\JJ^\C_2$ to $\JJ^\R_2$.
We have shown that on the spinorial side 
of the Newton-Hooke correspondence (\ref{NH}), a 4D Dirac spinor is reduced to the Majorana spinor with 4 real independent components (or alternatively Weyl spinor with 2 complex), that is, we impose
the  natural reality condition for spinors.

We have also noted that the Levi-Civita transform
connecting  the Landau problem and the 2D hydrogen atom 
is an extension of  the Hopf fibration  $S^1 \rightarrow S^1$. We have drawn the parallel with the Kustaanheimo-Stiefel transform thought as an extension of 
the Hopf fibration  $S^3 \rightarrow S^2$.
We believe that  phase space extentions of
the remaining  Hopf fibrations 
\[
\ba{ccccccccccc}
\H &&
0&\rightarrow &S^3 &\hookrightarrow& S^7 & \rightarrow &S^4 &\rightarrow& 0 \ , \\
\O&&
0&\rightarrow &S^7 &\hookrightarrow& S^{15} & \rightarrow &S^8 &\rightarrow& 0 \ ,
\ea
\]
are of potential interest for  high energy physics
in view of the exceptional role that the octonions could
play in the quark-lepton symmetry of the Standard Model \cite{DV,GG}.

\ack
It is a pleasure to thank Michel Dubois-Violette, Ludmil Hadjiivanov, Malte Henkel, Richard Kerner, Mariana Kirchbach, Petko Nikolov, Stoimen Stoimenov and  Ivan Todorov for many inspiring discussions. 
One of us (T.P.)  is supported 
by the Bulgarian National Science Fund Research Grant DN 18/3.

\end{document}